\documentclass[longauth]{aa} 

\usepackage{graphicx}
\usepackage{txfonts}
\usepackage{lscape}
\begin{document} 

\title{Very High Energy outburst of Markarian 501 in May 2009}
\titlerunning{VHE flare of Mrk\,501 in May 2009}

\author{
E.~Aliu\inst{1},
S.~Archambault\inst{2},
A.~Archer\inst{3},
T.~Arlen\inst{4},
T.~Aune\inst{4},
A.~Barnacka\inst{5},
B.~Behera\inst{6},
M.~Beilicke\inst{3},
W.~Benbow\inst{7},
K.~Berger\inst{8}
% ...
\\
(The VERITAS Collaboration)\\
M.~G.~Baring\inst{35},
J.~Becerra G{\'o}nzalez\inst{29,36},
A.~N.~Cillis\inst{30,29},
D.~Horan\inst{37},
D.~Paneque\inst{38,39}
}
\authorrunning{Aliu, Archambault, Archer et al.}

\institute{Department of Physics and Astronomy, Barnard College, Columbia University, NY 10027, USA\\
\and Physics Department, McGill University, Montreal, QC H3A 2T8, Canada\\
\and Department of Physics, Washington University, St. Louis, MO 63130, USA\\
\and Department of Physics and Astronomy, University of California, Los Angeles, CA 90095, USA\\
\and Harvard-Smithsonian Center for Astrophysics, 60 Garden Street, Cambridge, MA 02138, USA\\
\and DESY, Platanenallee 6, 15738 Zeuthen, Germany\\
\and Fred Lawrence Whipple Observatory, Harvard-Smithsonian Center for Astrophysics, Amado, AZ 85645, USA\\
\and Department of Physics and Astronomy and the Bartol Research Institute, University of Delaware, Newark, DE 19716, USA\\
\and School of Physics, University College Dublin, Belfield, Dublin 4, Ireland\\
\and Centre for Space Research, North-West University, Private Bag X6001, Potchefstroom 2520, South Africa\\
\and Santa Cruz Institute for Particle Physics and Department of Physics, University of California, Santa Cruz, CA 95064, USA\\
\and Department of Physics and Astronomy, Iowa State University, Ames, IA 50011, USA\\
\and School of Physics, National University of Ireland Galway, University Road, Galway, Ireland\\
\and Institute of Physics and Astronomy, University of Potsdam, 14476 Potsdam-Golm, Germany\\
\and Astronomy Department, Adler Planetarium and Astronomy Museum, Chicago, IL 60605, USA\\
\and Department of Physics and Astronomy, Purdue University, West Lafayette, IN 47907, USA\\
\and School of Physics and Astronomy, University of Minnesota, Minneapolis, MN 55455, USA\\
\and Department of Astronomy and Astrophysics, 525 Davey Lab, Pennsylvania State University, University Park, PA 16802, USA\\
\and Department of Physics, California State University - East Bay, Hayward, CA 94542, USA\\
\and Department of Physics and Astronomy, University of Iowa, Van Allen Hall, Iowa City, IA 52242, USA\\
\and Physics Department, Columbia University, New York, NY 10027, USA\\
\and Department of Physics and Astronomy, University of Utah, Salt Lake City, UT 84112, USA\\
\and Department of Physics and Astronomy, DePauw University, Greencastle, IN 46135-0037, USA\\
\and School of Physics and Center for Relativistic Astrophysics, Georgia Institute of Technology, 837 State Street NW, Atlanta, GA 30332-0430\\
\and Department of Physics, Anderson University, 1100 East 5th Street, Anderson, IN 46012\\
\and Department of Life and Physical Sciences, Galway-Mayo Institute of Technology, Dublin Road, Galway, Ireland\\
\and Department of Physics, Grand Valley State University, Allendale, MI 49401, USA\\
\and Enrico Fermi Institute, University of Chicago, Chicago, IL 60637, USA\\
\and N.A.S.A./Goddard Space-Flight Center, Code 661, Greenbelt, MD 20771, USA\\
\and Instituto de Astronomia y Fisica del Espacio, Casilla de Correo 67 - Sucursal 28, (C1428ZAA) Ciudad Autónoma de Buenos Aires, Argentina\\
\and Physics Department, California Polytechnic State University, San Luis Obispo, CA 94307, USA\\
\and Department of Physical Sciences, Cork Institute of Technology, Bishopstown, Cork, Ireland\\
\and Argonne National Laboratory, 9700 S. Cass Avenue, Argonne, IL 60439, USA\\
\and University of Maryland, College Park / NASA GSFC, College Park, MD 20742, USA\\
\and Rice University, Department of Physics and Astronomy, P.O. Box 1892, Houston, TX 77251-1892, USA\\
\and Department of Physics and Department of Astronomy, University of Maryland, College Park, MD 20742, USA\\
\and Laboratoire Leprince-Ringuet, École Polytechnique, CNRS/IN2P3, Palaiseau, France\\
\and W. W. Hansen Experimental Physics Laboratory, Kavli Institute for Particle Astrophysics and Cosmology, Department of Physics and SLAC National Accelerator\\
\and Max-Planck-Institut für Physik, D-80805 München, Germany
}

\date{Received ... ; accepted ...}

\abstract{The very high energy (VHE; {\it E} $>$ 100\,GeV) blazar Markarian\,501 was observed between April 17 and May 5 (MJD 54938--54956), 2009, as part of an extensive multi-wavelength campaign from radio to VHE. 
Strong VHE $\gamma$-ray activity was detected on May 1st with Whipple and VERITAS, when the flux ({\it E} $>$ 400\,GeV) increased to 10 times the pre-flare baseline flux ($3.9{\times 10^{-11}}~{\rm ph~cm^{-2}~s^{-1}}$), reaching five times the flux of the Crab Nebula. This coincided with a decrease in the optical polarization and a rotation of the polarization angle by 15$^{\circ}$. This VHE flare showed a fast flux variation with an increase of a factor $\sim$4 in 25 minutes, and a falling time of $\sim$50 minutes. 
We present the observations of the quiescent state previous to the flare and of the high state after the flare, focusing on the flux and spectral variability from Whipple, VERITAS, {\it Fermi}-LAT, RXTE, and {\it Swift} combined with optical and radio data. }

\maketitle
%

\section{Introduction}

Blazars are a subclass of active galactic nuclei (AGN) with relativistic jets
pointing along the line of sight to the observer. Blazars exhibit strong, rapid, and irregularly variable nonthermal emission over the entire electromagnetic spectrum, from radio to very-high-energy (VHE; {\it E} $>$ 100 GeV) $\gamma$ rays. Episodes of dramatic variability are produced in compact zones of the system, most likely in the relativistic jet (e.g. \citealt{Giannios09}). 
Blazars provide a unique opportunity to investigate this variability because they allow us to observe the processes occurring within the jets. This enables us to make inferences about the nature of the particles and the acceleration mechanisms that may be involved.

The blazar spectral energy distribution (SED) is characterized by a broad, double-peaked structure when plotted in $\nu F_{\nu}$, which is an indication of the broadband emission power. The peak at lower energies arises from synchrotron radiation from accelerated charged particles, while the second peak is explained by high energy processes of either leptonic (e.g. \citealt{MG85,maraschi92,dermer92,sikora94}) or hadronic (e.g. \citealt{aharonian00,DL97,BB99,PS00}) nature. Both peaks are found to vary with blazar activity.

Blazars exhibit outbursts in the optical, X-ray and VHE $\gamma$-ray bands. These flares have been observed to occur over various time scales, ranging from months to minutes. 
There are several plausible scenarios to explain the origin of the observed flares. For example, they can be caused by internal shock waves within the jet \citep{rees78,spada01} or the ejection of relativistic plasma \citep{bott97,MK97}. It has also been suggested that flares can be associated with magnetic reconnection events in a jet that is dominated by the magnetic field \citep{Lyu03}.
In some blazars, a strong correlation between X-ray and VHE $\gamma$-ray emission has been observed. This would imply that the same population of electrons is responsible for producing emission in both energy bands, via synchrotron and inverse-Compton emission (e.g. \citealt{coppi99,kraw00}).

To understand the actual mechanism and physical processes responsible for these emissions, it is essential to have long-term, 
well-sampled observations of a blazar across multiple energy bands (e.g. \citealt{steele07}). Multi-wavelength observations of $\gamma$-ray emitting blazars are thus an important tool for testing models of non-thermal emission from these objects. Measurements of the temporal correlation among flux variations at different wavelengths during flares are particularly useful and provide constraints on the emission models in various energy regimes (e.g. \citealt{ale15}).

Markarian 501 (Mrk\,501) is a member of the BL Lac subclass of blazars with a redshift of $\it{z}~=~$0.034. It was first detected as a VHE source by the Whipple 10\,m $\gamma$-ray telescope (hereafter Whipple) in 1996 \citep{quinn96}. Some observations of Mrk\,501 revealed a very low flux of  VHE $\gamma$ rays above 300 GeV at the level of about one tenth the flux of the Crab Nebula (e.g. \citealt{aha05,alb08,ale15}). In 1997, however, Mrk\,501 exhibited an unprecedented flare in VHE $\gamma$ rays with an integral flux of up to four times the flux of the Crab Nebula \citep{cat97,pian98,petry00}. The shortest flux variability measured in Mrk\,501 has a rise/fall time of a few minutes \citep{alb07}. Even though Mrk\,501 is a highly variable source of VHE $\gamma$-ray emission, it has shown fewer flares and changes in the flux activity than Markarian\,421, the first-discovered \citep{Punch92} and well-studied extragalactic VHE $\gamma$-ray source \footnote{A list of some of the papers describing observations of Mrk\,421 is available here: \url{http://tevcat.uchicago.edu/?mode=1&showsrc=75 }}.

Mrk\,501 has been the target of many multi-wavelength campaigns mainly covering VHE flaring activity with intra-night variability for a few days (e.g. \citealt{cat97,alb07}) and over several months independently of the source activity (e.g. \citealt{kranich09,pichel09}). 
As an example of long and short-term variability, \citealt{quinn99} presented observations over four years (1995-1999) during which they detected significant variability in the monthly average flux and also rapid VHE flares lasting for a few hours on individual nights.

 As part of a large-scale multi-wavelength campaign over a period of 4.5 months in 2009, 
Mrk\,501 was observed from April 17 to May 5, 2009 (MJD 54938--54956) with a number of ground- and space-based observatories covering the spectrum from radio to VHE $\gamma$ rays and including optical polarization. The average SED of Mrk\,501 for this campaign is well described by the standard one-zone synchrotron self-Compton (SSC) model \citep{abdo10}. In this paper we report on the observations taken in this period and particularly on the flare of May 1 detected by Whipple, when the source flux rose to approximately five times the flux of the Crab Nebula ($\sim$ 50 times the integral flux detected with Whipple in 1996). 
The study which relates the multi-band variability and correlations using the full data set from the entire multi-instrument campaign will be reported in \citet{ahnen}.

\section{Data set and data reduction}

Several observatories participated in the 3-week multi-wavelength campaign reported in this paper. Table \ref{mrk501_0809} summarizes the data set for each instrument. Comprehensive coverage of the electromagnetic spectrum from radio to VHE $\gamma$ rays was achieved during the campaign, as shown in Figures \ref{double_fig} and \ref{m5pol}. The X-ray and $\gamma$-ray bands were well-sampled, including some simultaneous observations. In this section, we describe the observations taken in each waveband: VHE $\gamma$ rays with Whipple and VERITAS (Section \ref{vhe}) \footnote{MAGIC also participated in the overall multi-wavelength campaign, but could not observe Mrk501 during the 3-week period considered in this paper owing to bad weather and a hardware system upgrade occurring during the period MJD 54948--54960 (April 27 - May 13).}; high-energy (HE; 20 MeV -- 300 GeV) $\gamma$ rays with {\it Fermi}-LAT (Section \ref{he}); X-rays with {\it Swift}-XRT and RXTE (Section \ref{xr}); optical with GASP, MitSume, {\it Swift}-UVOT and Steward Observatory (Section \ref{optical}); and radio with Mets\"ahovi and OVRO (Section \ref{radio}).

\begin{table*}
\caption{Data set of Markarian 501 for the 3-week multi-wavelength campaign in 2009. 
Each data set was fitted with a constant flux model and the goodness of the fit test is shown in column 5.}
\label{mrk501_0809}      
\centering    
\begin{tabular}{c c c c c }
\hline
\hline
Waveband & Instrument & MJD Range & $\chi^{2}$/NDF \\
\hline 
VHE $\gamma$-ray &  Whipple & 54938-54955  & 279.3/16 \\
 & VERITAS & 54938-54955  & 184.7/5 \\
\hline
HE $\gamma$-ray & {\it Fermi}-LAT & 54938-54956  & 2.2/4 \\ 
\hline
X-ray & {\it Swift}-XRT low & 54941-54955  & 84.0/7 \\
 & {\it Swift}-XRT high & 54941-54955  & 98.0/7 \\
 & RXTE-PCA & 54941-54956  & 11.9/3 \\
\hline
Optical & GASP R & 54938-54955  & 16.6/9 \\
 & MitSume g & 54948-54956  & 3.1/3 \\
 & {\it Swift}-UVOT & 54941-54955  & 50.1/9 \\
 & Steward Observatory & 54947-54955  & 234.2/7 \\
\hline
Radio & Mets\"ahovi 37 GHz & 54942-54956  & 15.8/11 \\
 & OVRO 15 GHz & 54940-54955  & 4.6/4   \\
\hline
\end{tabular}
\end{table*}

\subsection{VHE $\gamma$-ray observations: Whipple/VERITAS}
\label{vhe}

Whipple \citep{kildea07} was located at the Fred Lawrence Whipple Observatory (FLWO), in southern Arizona, at an elevation of 2 312\,m above sea level. 
The telescope was built in 1968 and comprised a 10-meter optical reflector, composed of 248 spherical front-aluminized glass mirrors in a Davies-Cotton design \citep{weekes72,DV57}. The camera, located in the focal plane, was upgraded in 1999 \citep{finley99} to 379 photomultiplier tube (PMT) pixels sensitive in the ultraviolet (UV), with a quantum efficiency of $\sim$20\,$\%$. Each PMT had a $0.12^{\circ}$ field of view (FOV), giving a total FOV of $2.6^{\circ}$ for the camera. The telescope was sensitive in the energy range from 200 GeV to 20 TeV, with a peak response energy (for a Crab-like spectrum; power-law with $\Gamma$=2.6) of approximately 400 GeV during the observations presented here when analyzed with the standard analysis parameters, described in \citet{victorphd}.
From 2005 to 2012, the Whipple observing plan focused on the monitoring of VHE $\gamma$-ray-bright blazars, including Mrk\,501. As an example, the VERITAS observations of the VHE flare and the following nights presented in this paper are a direct result of this monitoring initiative. 

Whipple observed in two different modes, ON/OFF and TRK (tracking). 
For the ON and TRK runs, the source was centered on the target and the telescope tracked it for 28 minutes. For the background estimation, the OFF run was collected at an offset of 30 minutes, both in time and in right ascension, also for a duration of 28 minutes. In this way, the ON and OFF runs were taken at the same declination over the same range of telescope azimuth and elevation angles. This removes systematic errors that depend on slow changes in the atmosphere. In the TRK mode, there were no separate OFF observations. The background was instead estimated from events that passed all of the gamma-ray selection criteria except for the orientation cuts \citep{cat98}.

Whipple observed Mrk\,501 every night from April 17 to May 5, 2009, for a total of 20 hours of live time, with an overall detection of 11\,$\sigma$ and a mean flux corresponding to 30\,$\%$ of the Crab Nebula. 
To provide a comparison between the results that were obtained by Whipple and VERITAS, the VHE light curve is shown with a common energy threshold of 300\,GeV (see Figure 1). 
To do this, a power-law spectrum with index 2.5 (similar to the mean index found for the source) was used to normalize the integral flux of the Whipple data (with an energy threshold of $\sim$400\,GeV) to an integral flux above 300\,GeV.\\

VERITAS is an array of four atmospheric Cherenkov telescopes located at the basecamp of the FLWO in southern Arizona, at an altitude of 1 268\,m above sea level \citep{holder06}. 
During the time of the reported observations, VERITAS was sensitive in the energy range from 100\,GeV to 30\,TeV. The telescope design is based on Whipple, with each of the four telescopes consisting of a 12\,m diameter segmented reflector with a Davies-Cotton design supporting 354 hexagonal mirror facets. Each camera comprises 499 PMTs that have individual FOVs of $0.15^{\circ}$, which combine to give a total camera FOV of $3.5^{\circ}$ at the focus.

The VERITAS sensitivity has improved over the years owing to developments in data analysis techniques, optical alignment, calibration and, most significantly, by the relocation of the original prototype telescope (now Telescope 1) in 2009 after these data were taken and the PMT upgrades in 2012.
These upgrade occurred after the acquisition of the data presented here. The original array could detect a 1\,$\%$ Crab Nebula flux source in approximately 50 hours of observations (assuming a Crab Nebula spectral shape, \citealt{ong09}). This can be achieved in half of that time post upgrade \citep{park15}.

Observations are performed using the so-called wobble mode of operation, in which all telescopes are pointed with 0.5$^{\circ}$ offset in each of 4 directions with respect to the source position. 
This method allows for simultaneous estimates of the source and background flux \citep{fomin94}.

VERITAS took observations on Mrk\,501 during the reported period for four hours. 
Owing to the relocation of Telescope 1 and a temporary hardware issue only two or three telescopes from the full telescope array were operational during these observations: two telescopes for the nights of April 30 and May 1, and three telescopes for the rest of the nights. The overall detection was at a level of 34 $\sigma$ with an energy threshold of 300\,GeV.

\subsection{HE $\gamma$-ray observations: {\it Fermi}-LAT}
\label{he}

The Large Area Telescope (LAT) on board the {\it Fermi} Gamma-ray Space Telescope satellite is designed to observe electromagnetic radiation in the 20\,MeV to more than 300\,GeV energy band. {\it Fermi}-LAT has a peak effective area of 0.7\,${\rm m^{2}}$ for 1\,GeV photons, an energy resolution typically better than 10\,$\%$ and a FOV of about 2.4\,sr (20\,$\%$ of the entire sky), with an angular resolution (68\,$\%$ containment angle) better than ${\rm 1}^{\circ}$ for energies above 1\,GeV. Further details on the LAT can be found in \citet{lab11,ack2012}.

The analysis was performed with the ScienceTools software package version v9r33p0, which is available from the
Fermi Science Support Center \footnote{\url{http://fermi.gsfc.nasa.gov/ssc/}}. 
The Pass7 reprocessed SOURCE class events were extracted from a circular region of 10$^{\circ}$ radius centered at the location of Mrk\,501. The analysis was performed using photon energies greater than 0.3 GeV to be less sensitive to possible contamination from neighboring sources. A cut on the zenith angle ($<$ 100$^{\circ}$) was also applied to reduce contamination by $\gamma$ rays from the Earth limb, which are produced by cosmic rays interacting with the upper atmosphere. The background model used to extract the $\gamma$-ray signal comprises a Galactic diffuse-emission component ({\it gll$\_$iem$\_$v05$\_$rev1}) and an isotropic component ({\it iso$\_$source$\_$v05}). The normalizations of both components in the background model were allowed to vary freely during the spectral fitting. In addition, all 2FGL (Second Fermi Catalog; \citealt{2docat}) sources within 15$^{\circ}$ of Mrk\,501 were included.
The spectral analysis was performed with the post-launch instrument response functions P7REP$\_$SOURCE$\_$V15 using a binned maximum-likelihood method.
In the source model, the parameters of all point sources with a distance $<$10$^{\circ}$ from the center of the region of interest (ROI) were allowed to vary freely. For sources at $>$10$^{\circ}$, the normalization and the photon index were fixed to their values from the 2FGL catalogue.
The systematic uncertainties are dominated by the uncertainties on the effective area, 
and are estimated to be between 5 and 10\% in the energy range 100\,MeV to 100\,GeV. For more information regarding these uncertainties, see \citet{ack2012}.

{\it Fermi}-LAT operates in survey mode, which means that any point of the sky is observed for 30 min approximately every three hours. However, as Mrk\,501 is a relatively weak source for {\it Fermi}-LAT, an integration over several days is typically required to obtain a significant detection although sometimes it can be detected in daily average \footnote{http://fermi.gsfc.nasa.gov/ssc/data/access/lat/msl$\_$lc/source/Mrk$\_$501}.

\subsection{X-ray observations}
\label{xr}

Mrk\,501 was observed by {\it Swift} in 2009 as part of a long-term monitoring campaign, with increased coverage in April-May 2009 that comprised ten observations in the period MJD\,54941--54955. The X-ray telescope (XRT) on board the {\it Swift} satellite \citep{gehrels04} is sensitive in the 0.2-10\,keV energy range. The {\it Swift}-XRT data were analyzed using the HEASOFT package (version 6.11). The data were taken in the window-timing (WT) and photon-counting (PC) modes. The events were selected from grades 0 to 2 for WT mode and 0 to 12 for PC mode, over the energy range 0.3-10\,keV \citep{burr05}. 
Source counts were extracted from a rectangular region of 40 pixels long by 20 pixels wide centered on the source. For PC mode data, events were selected within a circle of 20 pixel ($\sim$46 arcsec) radius, which encloses about 80\% of the point spread function (PSF), centered on the source position.
Background counts were extracted from a nearby source-free rectangular region of equivalent size. Ancillary response files were generated using the {\it xrtmkarf} task, with corrections applied for the PSF losses and CCD defects. The corresponding response matrix from the XRT calibration files (CALDB tag v.011) was applied. 

The 0.3-10 keV source energy spectra were binned to have more than 20 counts per bin before the spectral fitting was performed. The spectra were corrected for absorption with a neutral hydrogen column density fixed to the Galactic 21 cm value in the direction of Mrk\,501 (1.56 $\times 10^{20} {\rm cm^{-2}}$; \citealt{kalberla}).

The X-ray satellite mission RXTE \citep{bradt93} observed Mrk\,501 in four exposures in the period MJD 54941--54956. The Proportional Counter Array (PCA) instrument \citep{jah96} is comprised of five proportional counter units (PCUs) covering a nominal energy range of 2-60\,keV. Data reduction was performed with the HEASOFT package (version 6.11). Only the top layer (X1L and X1R) signal was used. The data were filtered following the standard criteria advised by the NASA Guest Observer Facility\footnote{http://heasarc.nasa.gov/docs/xte/xhp$\_$proc$\_$analysis.html}. 
Background data were parameterized with the {\it pcabackest} tool using the {\it pca$\_$bkgd$\_$cmfaintl7$\_$eMv20051128.mdl$\_$pca$\_$saa$\_$history}.gz model for faint sources. The photon spectrum of each observation was extracted using the {\it saextrct} tool. Response matrices were generated using {\it pcarsp} with the calibration files. For further details on the analysis of faint sources with RXTE, see the online cook book (\url{http://heasarc.gsfc.nasa.gov/docs/xte/recipes/cook_book.html}).

\subsection{Optical observations}
\label{optical}

The optical fluxes reported in this paper were obtained within the GASP-WEBT program \citep[e.g.][]{Villata2008, Villata2009}, with various optical telescopes around the globe, and by the two MitSume telescopes, which are located in Yamanashi and Okayama (Japan). Optical polarization measurements are also included from the Steward Observatory.

The fluxes from GASP were obtained with the R filter, while the ones from MitSume were obtained only with the $g$ filter. The complete set of optical data will be presented in \citet{ahnen}. These instruments used the calibration stars reported in \citet{Villata1998}, and the Galactic extinction was corrected with the coefficients given in \citet{schlegel98}. The flux from the host galaxy, which in the $R$ band accounts for about two-thirds of the overall measured optical flux \citep[]{Nilsson2007}, was not subtracted. As can be seen below (Section \ref{SED}), in the SED of Figure \ref{m5_sed}, the host-galaxy contribution shows up as an additional narrow bump with the peak located at infrared frequencies and the flux decreasing rapidly with increasing frequency. 

The UV data points reported here were obtained with the Swift-Ultra-Violet/Optical Telescope \citep[UVOT;][]{Roming2005}. Three UV colors from UVOT were used, namely the $W1$, $M2$, and $W2$ filters.
Photometry was computed using a $5$\,arcsec source region around Mrk\,501 applying a custom UVOT pipeline (FTOOLS version 6.7). This pipeline was validated with the public pipeline reported in \citet{Poole2008}. The advantage of the custom pipeline is that it allows for separate observation-by-observation corrections for astrometric misalignments, as reported in \citet[]{AcciariMrk4212008}.  A visual inspection was also performed on each of the observations to ensure proper data-quality selection and correction.
The flux measurements obtained have been corrected for Galactic extinction $E_{B-V} = 0.02$\,mag \citep[]{schlegel98} in each spectral band \citep[]{Fitzpatrick99}. See Table \ref{mrk501_0809} for details on the time interval and the number of observations performed with all these instruments.

Optical flux and polarization observations during the high-energy monitoring campaign were obtained using the 2.3\,m Bok Telescope of Steward Observatory (SO), located on Kitt Peak, AZ.  These observations are part of the public SO program to monitor gamma-ray-bright blazars during the Fermi-LAT mission\footnote{\url{http://james.as.arizona.edu/~psmith/Fermi}}  \citep{SOref2}.  Mrk 501 was observed on each night from MJD 54947 to MJD 54955, which included the night of the VHE flare, using the SPOL imaging/spectropolarimeter \citep{SOref1}.

Uncertainties in the degree of linear polarization (P) and the electric vector position angle (EVPA) of the polarized flux are about 0.05\% and 0.3$^{\circ}$ respectively. These uncertainties are completely dominated by photon statistics because known sources of systematic errors are effectively eliminated due to the dual-beam design of SPOL and the fact that the data were obtained over a full rotation (16 positions) of the wave plate.

\subsection{Radio observations}
\label{radio}

The radio data reported here were taken with the Owens Valley Radio Observatory (OVRO) 40\,m telescope, observing at a frequency of 15 GHz, and the 14\,m Mets\"ahovi radio telescope observing at 37 GHz. The data were reduced
according to the prescription given in \citet[OVRO]{Richards2011} and \citet[Mets\"ahovi]{Terasranta1998}.
For these two single-dish telescopes, Mrk\,501 is a point-like and unresolved source, and hence 
the flux reported denotes the total flux density integrated over the source. 
Consequently, these fluxes were taken as upper limits in the SED model fit shown in Section \ref{SED}.

\section{Light curves}
\label{lc}
\begin{figure*}
\centering
\includegraphics[width=0.88\textwidth]{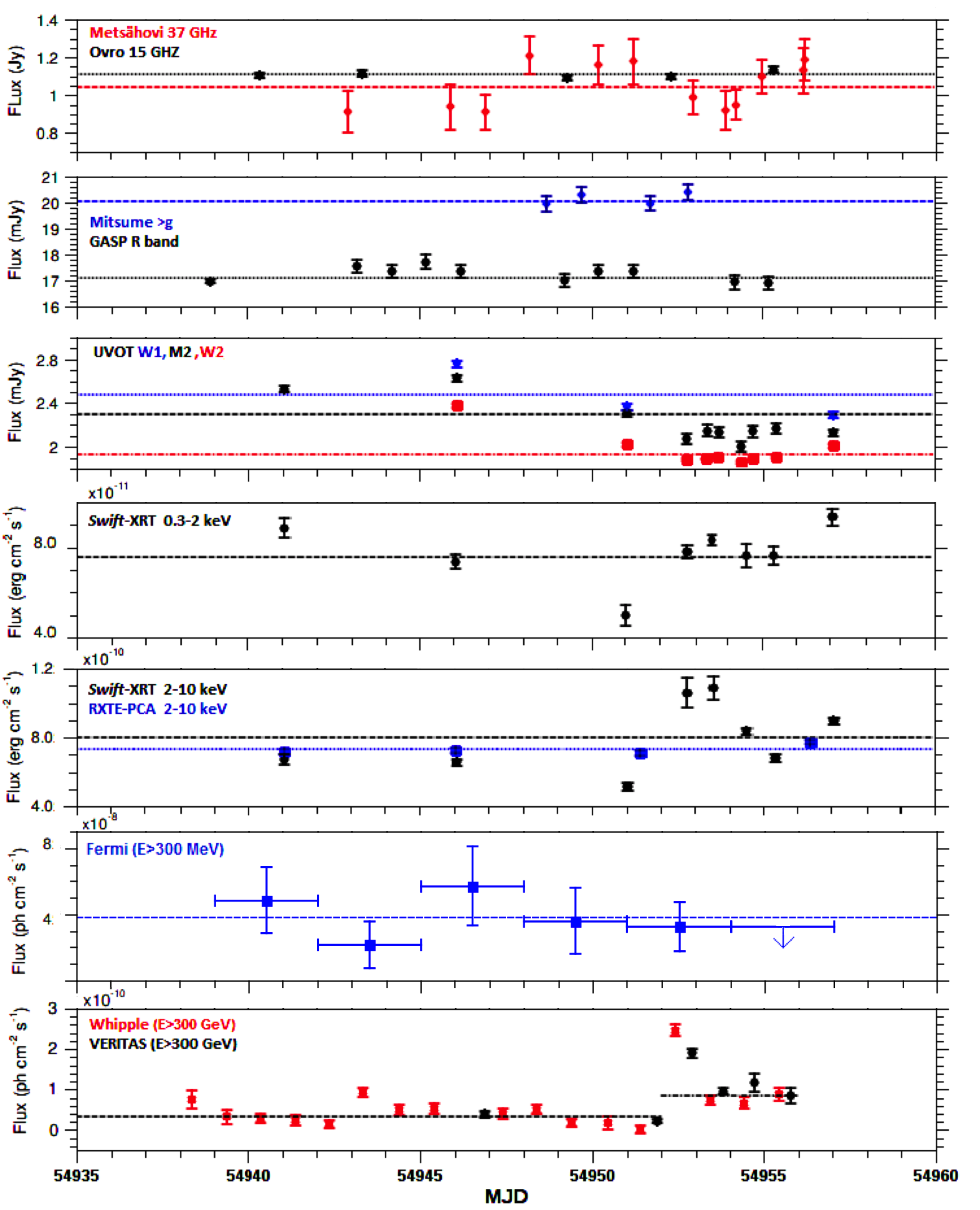}
   \caption{Daily average light curves for Mrk 501 from April 17 to May 5, 2009. Each dotted horizontal line represents a constant line fit for each instrument involved. {\sl Top}: OVRO at 15 GHz (black filled circles) and Mets\"ahovi at 37 GHz (red filled diamonds); {\sl Second}: Mitsume in g band (blue filled diamonds) and GASP in R band (black filled circles). {\sl Third}: {\it Swift}-UVOT in the ultraviolet, with three different bands, UVW1 (260nm, blue diamonds), UVM2 (220nm, black circles) and UVW2 (193nm, red squares). {\sl Fourth}: X-ray: {\it Swift}-XRT 0.3-2 keV. {\sl Fifth}: X-ray: RXTE-PCA (blue squares) and {\it Swift}-XRT (black circles) 2-10 keV (nightly average); {\sl Sixth}: HE gamma-ray: {\it Fermi}-LAT (E $>$ 300 MeV; 3-day average); {\sl Bottom}: VHE $\gamma$-rays: Whipple ({\it E} $>$ 400 GeV, normalized to {\it E} $>$ 300 GeV; red filled stars) and VERITAS ({\it E} $>$ 300 GeV; black filled circles). The different dotted lines are the constant fit for the low and high state.
   }
  \label{double_fig}
\end{figure*}

\begin{figure*}
\centering   
  \includegraphics[width=0.88\textwidth]{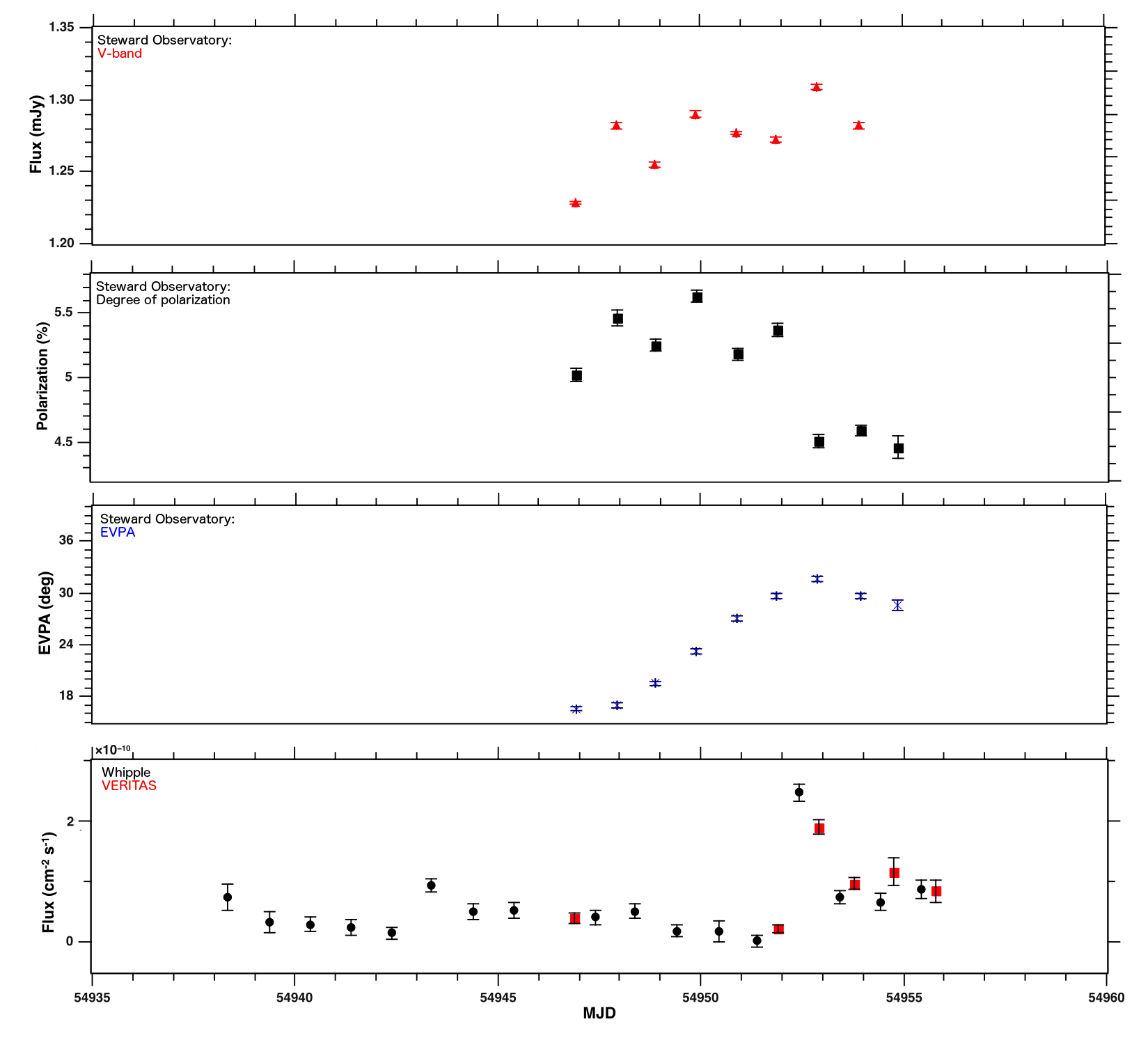}
  \caption{\footnotesize{Optical flux in the V band, degree of the optical linear polarization and electric-vector position angle versus time (first, second and third plots respectively) measured at the Steward Observatory, and (fourth plot) the VHE light curve obtained with Whipple and VERITAS.}}
  \label{m5pol}
 \end{figure*}

The light curves from all of the observations taken on Mrk\,501 as part of the multi-wavelength campaign
from April 17 to May 5, 2009 (MJD 54938--56) are shown in Figure \ref{double_fig}, except for the optical observations taken
with the Steward telescopes, which are considered separately at the end of this section (see Figure \ref{m5pol}).  We fitted a constant to all the light curves and we show the corresponding  $\chi^{2}$ values in Table \ref{mrk501_0809}.

In the radio and optical bands (except the observations taken by the Steward Observatory that shows high variability), the measured fluxes were constant (within statistical uncertainties). The UV band shows some variations (around 20$\%$), although during the VHE flare the flux was steady.
At X-ray energies, the light curves show some variation around the epoch of the VHE flare, up to a factor of two.
In the VHE domain, VERITAS, and especially Whipple, measured statistically significant flux variations
of a factor of a few and up to a factor of ten for MJD 54952. During the mentioned 3-week
time interval, the highest variability was found to be at the highest energies.

At VHE, the light curve is consistent with constant emission (above 300\,GeV) by the source ($3.9{\times 10^{-11}}~{\rm ph~cm^{-2}~s^{-1}}$; hereafter baseline emission) until the night of May 1 (MJD 54952), when a high-emission state was detected first with Whipple and 1.5 hours later with VERITAS, reaching a maximum $\gamma$-ray flux of $\sim$10 times the average baseline flux, approximately five times the Crab Nebula flux. VERITAS continued with simultaneous observations with Whipple until the end of that night.

\begin{figure}
  \centering
 \includegraphics[width=\hsize]{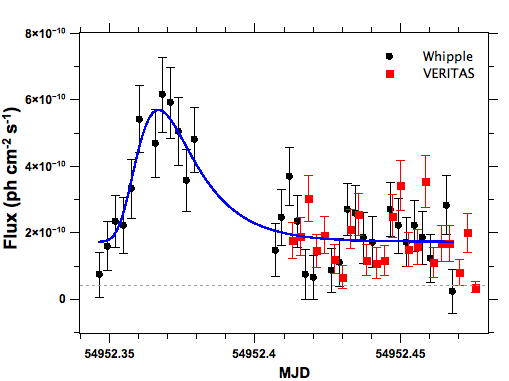}
  \caption{\footnotesize{Whipple and VERITAS (E $>$ 300\,GeV) light curve (4-minute binning) for the night of the VHE flare. The dotted line shows the baseline emission for the source based on the flux levels depicted in Figure \ref{double_fig}. Blue curve: VHE flare model fitted to the Whipple and VERITAS data}} 
  \label{fig2}
 \end{figure}

Figure \ref{fig2} shows the Whipple and VERITAS light curves in 4-minute bins for May 1, 2009 (MJD 54952), with the flux increasing by a factor of $\sim$4 in the first 25 minutes. During the days after the flare (MJD 54953--55), the source remained in an elevated state; the flux being about twice the baseline flux each night.

An approach to get characteristic parameters for the VHE flare has been performed with a very simple flare model
\citep{alb07}, in which the amplitude, duration, and rise/fall times  of the flare are quantified.
The model parameterizes a flux variation (flare) $F(t)$ superposed on a stable emission as
\begin{equation}\label{ecuaciona}
\displaystyle F(t)=~a~+~\frac{b}{2^{{(t-t_{0})/d}}~+~2^{-(t-t_{0})/c}}
\end{equation}

\noindent where {\it a} is the baseline emission after the flare; $t_{0}$ is the time when the highest flux in the light curve was observed; and {\it b}, {\it c}, and {\it d} are fit parameters. The {\it c} and {\it d} parameters denote the flux-doubling rise and fall times, respectively. 
The result of the fit is shown in Figure \ref{fig2}, in which the combined Whipple and VERITAS data were considered, and the best fit parameters are {\it a}=(18$\pm$2)$\times10^{-11}~{\rm ph~cm^{-2}~s^{-1}}$, {\it b}=(73$\pm$9)$\times10^{-11}~{\rm ph~cm^{-2}~s^{-1}}$, {\it c}=(1580$\pm$110)$~{\rm s}$ and {\it d}=(2920$\pm$240)$~{\rm s}$, yielding a ${\chi^{2}}$/dof of 6.4/31.

The values obtained for the fall and the rise time using the fit were done using a baseline emission after the flare ({\it a} parameter) four times higher than the baseline emission found at low state of activity, prior to the flare. 

The fall ($\sim$ 50\,minutes) and rise ($\sim$ 25\,minutes) times were both sufficiently short to imply that the emission region was very small, constrained by 
$R \leq ct_{var} \delta (1 + z)^{-1}$ , where $\delta$ is the relativistic Doppler factor. 
The variability timescale was not as fast as that observed in PKS\,2155$-$304 by \citet{Aharonian07}, in which the rise and decay times were of the order of two to three minutes. Also, Mrk\,501 showed this fast variability in previous observations \citep{alb07} with a rise and fall time of around three minutes.

In the X-ray bands, RXTE-PCA data taken in four exposures at 5-day intervals showed no statistically significant variations during this 3-week period. It is worthwhile noting that the closest  RXTE-PCA observations to the night of the VHE activity detected by Whipple were taken more than a day before and more than three days later. {\it Swift}-XRT showed variations in both bands, with a decrease of $\sim$20-30\% until $\sim$1 day prior to the VHE flare, followed by an increase of $\sim$70-100\% seven hours after the flare. 

The light curve with the data binned in 3-day time intervals for {\it Fermi}-LAT is also presented in
Figure \ref{double_fig}. The time interval containing the entire VHE flare (started on MJD 54952) does not show any significant variation with respect to the previous ones. However, the source had a flux enhancement by a factor of four compared to the average flux reported in the 2FGL.

 Optical polarization measurements taken at the Steward Observatory are shown in Figure \ref{m5pol}. In the figure, the optical flux, the degree of the optical linear polarization {\it P}, and the electric-vector position angle EVPA are shown for the nights corresponding to the VHE flare, together with the light curve at VHE for comparison. The polarization {\it P} was approximately steady at about 5\% for the period MJD 54947--51, and dropped from 5.3\% to 4.5\% after the VHE flare.
EVPA showed a continuous increase from $\sim$15 to $\sim$30 degrees in five days, decreasing a few degrees immediately after the large VHE flare occurred. We can compare these measurements with previous observations. Mrk 501 never exceeded {\it P} = 4.2\% in 38 observations that were obtained in 1987-1990 (\citealt{jannuzi94}) except at shorter wavelengths (U) not covered by the SPOL observations.  Also, the collation of optical polarization measurements of Mrk\,501 made prior to 1986 by \citet{rusk90} shows that the object varied from 2 to 4\% in P.  Similarly, the EVPA varied within a restricted range (125$^{\circ}$-145$^{\circ}$) that is near the position angle of the inner  jet as determined by VLBI observations (e.g. \citealt{rusk88}), and not radically different from the later observations of \citet{jannuzi93}, which showed the EVPA in the range 90$^{\circ}$-130$^{\circ}$.  
The EVPA was near one of its apparent limits (15$^{\circ}$-35$^{\circ}$) during the epoch surrounding the VHE outburst, starting with a lower value (1-2\%) the previous month (March 2009). 
As shown in Figure \ref{m5pol}, the EVPA reaches $\sim$30$^{\circ}$ at the peak of the $\gamma$-ray flare and reverses its direction of rotation as the outburst fades.  Both in terms of the high level of optical polarization and the polarization position angle, the VHE outburst in Mrk\,501 was accompanied by unusual polarization behavior. If these events are physically linked, this might indicate a common origin for the optical and $\gamma$-ray emission as has already been seen in other sources (\citealt{lab17,lab18,lab19}).

\section{Flux variability and correlation}

A correlation in the variability at different wavelengths can give indications about, or put constraints on, the processes involved in the emission mechanism. Although a correlation between the variability in the X-ray and VHE $\gamma$-ray fluxes has often been observed in Mrk\,501, it is not yet certain that this is always the case for this blazar.
A clear correlation was found in several studies for different flaring sources (e.g. \citealt{lab13,lab14}), and no correlation in some other different flaring sources (e.g. \citealt{kraw04}).
For this kind of study, simultaneity between observations at different bands is critical.

The first approach to search for variability is to establish whether there is intrinsic variability in a given band alone. In this work, the light curves were tested with a constant flux model and the $\chi^{2}$ results to check the consistency with that model, are shown in Table \ref{mrk501_0809}, as discussed above.
In order to go further in quantifying the flux variability present in the light curves, the fractional RMS variability amplitude, $F_{\rm var}$ \citep{edelson02,lab12}, is calculated as
\\
\begin{equation}\label{ecuacion1}
\displaystyle F_{{\rm var}}=\sqrt[]{\frac{S^{2}- \left<{\sigma}^{2}\right>} {{\left<F\right>}^{2}}}
\end{equation}

\noindent where $\left<F\right>$ is the average photon flux, $S$ is the standard deviation of the total $N$ flux measurements, and $\left<{\sigma}^{2}\right>$ is the mean squared error of those $N$ measurements, all for a given energy interval. $F_{{\rm var}}$ is commonly used to measure the variability (after subtracting the expected contribution from observation uncertainties) of a series of measurements, typically obtained during a campaign (e.g. \citealt{horan09,edelson02}). An $F_{{\rm var}}$ value close to zero indicates that there was no significant detectable variability over the period, and a value close to one indicates strong variability.

\begin{figure}
  \centering
 \includegraphics[width=\hsize]{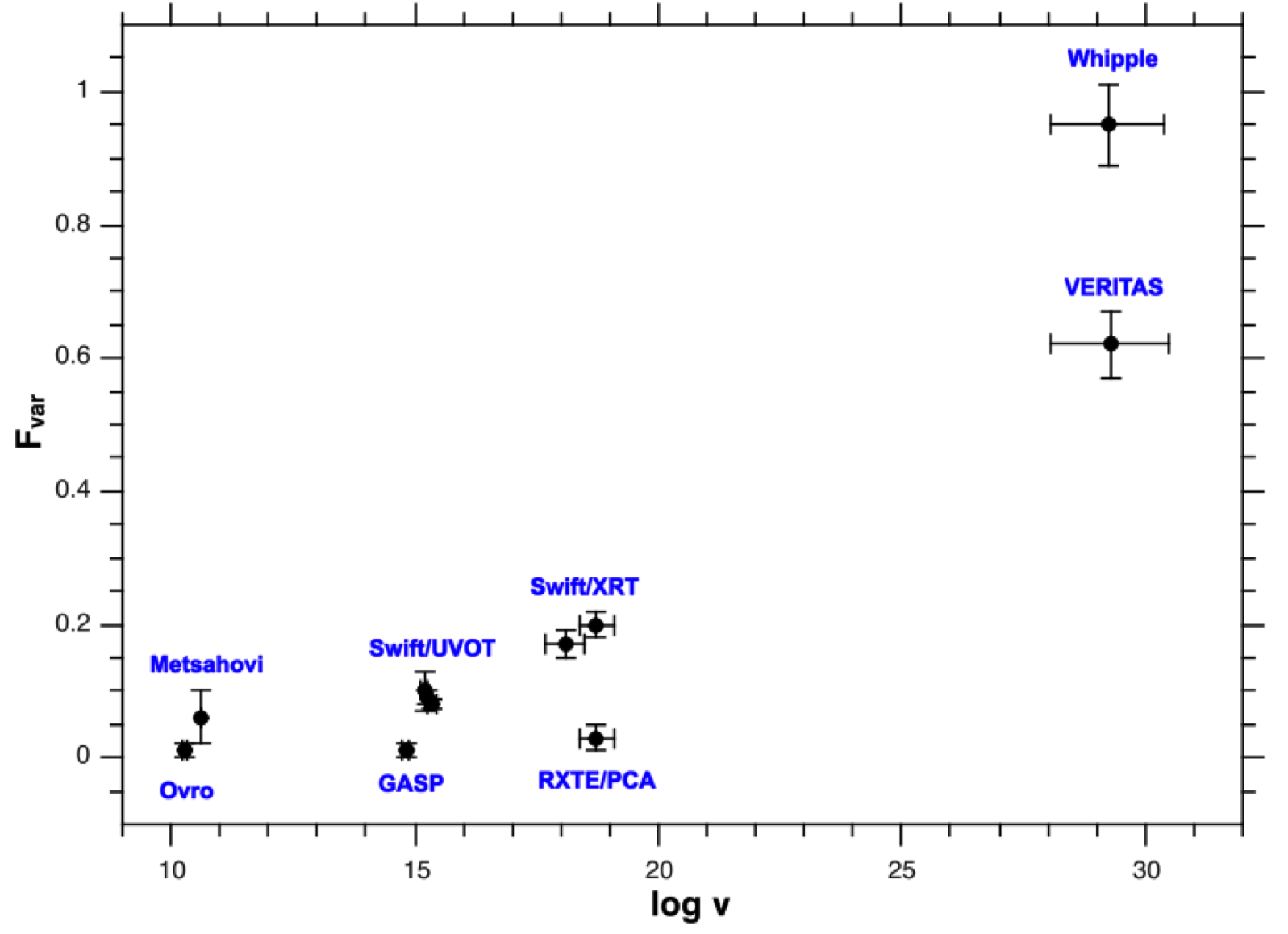}
  \caption{Multi-band fractional variability amplitude for Mrk\,501 during the 3-week period MJD 54938--54956.}
  \label{fig3}
 \end{figure}

Figure \ref{fig3} shows the $F_{\rm var}$ values obtained for all the energy bands involved using a daily average for each energy band. 
{\it Fermi}-LAT was excluded because it shows a negative excess ($\left<{\sigma}^{2}\right> > S^{2}$), thus indicating the errors are larger than any flux variations that might be present.
Essentially this type of result can be interpreted as null evidence for variability, because either there was no variability or, more likely, the instrument was not sensitive enough to detect it. \citet{abdo10} found a value between 0.3 and 0.4 for the $F_{{\rm var}}$ using 16 months of data.  
The value of the $F_{{\rm var}}$ for GASP R is substantially smaller than {\it Swift}-UVOT because of the contribution of the host galaxy, which is not subtracted and contributes with about 2/3 of the overall flux in the R band (while it is essentially negligible in the UV band).
As seen in the figure, the values of $F_{{\rm var}}$ are either very low or compatible with zero for all of the energy bands, except for the two data sets in the VHE domain, where $F_{\rm var}$ is 0.6$\pm$0.1 for VERITAS and 0.9$\pm$0.1 for Whipple, and for the {\it Swift}-XRT, where a value of $F_{\rm var} \sim$ 0.20$\pm$0.02 is found for the data in each waveband. 
The large variability in the VHE domain is clearly dominated by the high VHE flare observed on MJD 54952 and the following few days.

\begin{figure}[!h]

\includegraphics[width=\hsize]{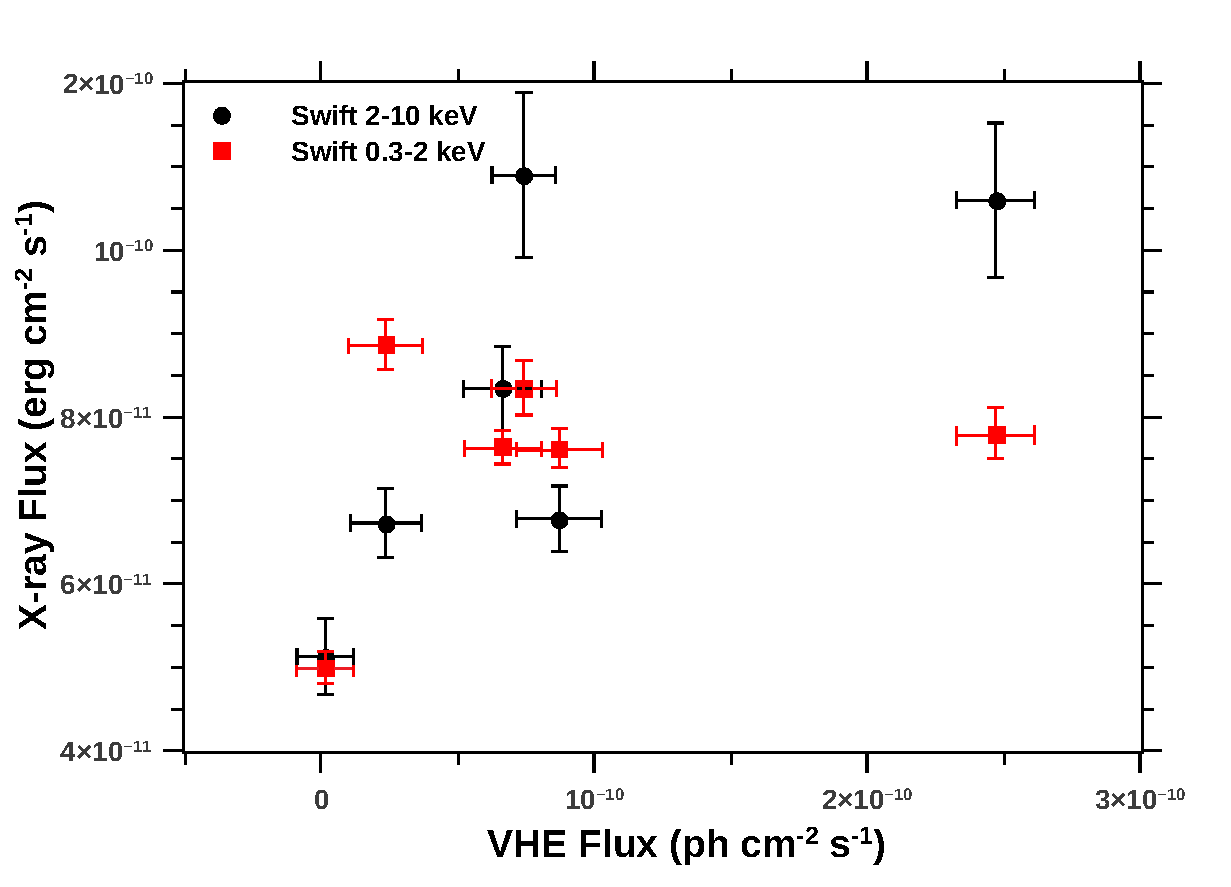}  
  \caption{Flux-flux correlation for X-rays and VHE $\gamma$ rays taken with Whipple. Only pairs of observations within 24 hours of each other were used.} 
  \label{fig4}
 \end{figure}

Having found intrinsic variability in both X-ray and VHE $\gamma$-ray bands, we study the flux correlation of simultaneous data. 
As strict simultaneity is difficult to achieve, this condition was relaxed to a window of 24\,hr duration. As this timescale is greater than the variability timescale of the flare, it is likely that data taken across different flux states are being combined, which may affect the correlation study.
The results are shown in Figure \ref{fig4} for Whipple (the most complete set and having the highest variability in the VHE $\gamma$-ray domain) and for the {\it Swift}-XRT for its two energy bands. Since the VHE flare had a timescale of $\sim$ 30 minutes, and there are not any X-ray data within a comparable time scale (the observations were taken seven hours before the VHE flare), it is not at all surprising that this data point does not contribute to the overall VHE-X-ray correlation.
Excluding the flare, there is an indication of a trend.
However, there are too few data points to make a claim about the apparent relation between these bands.

\begin{figure}
  \centering
  \includegraphics[width=\hsize]{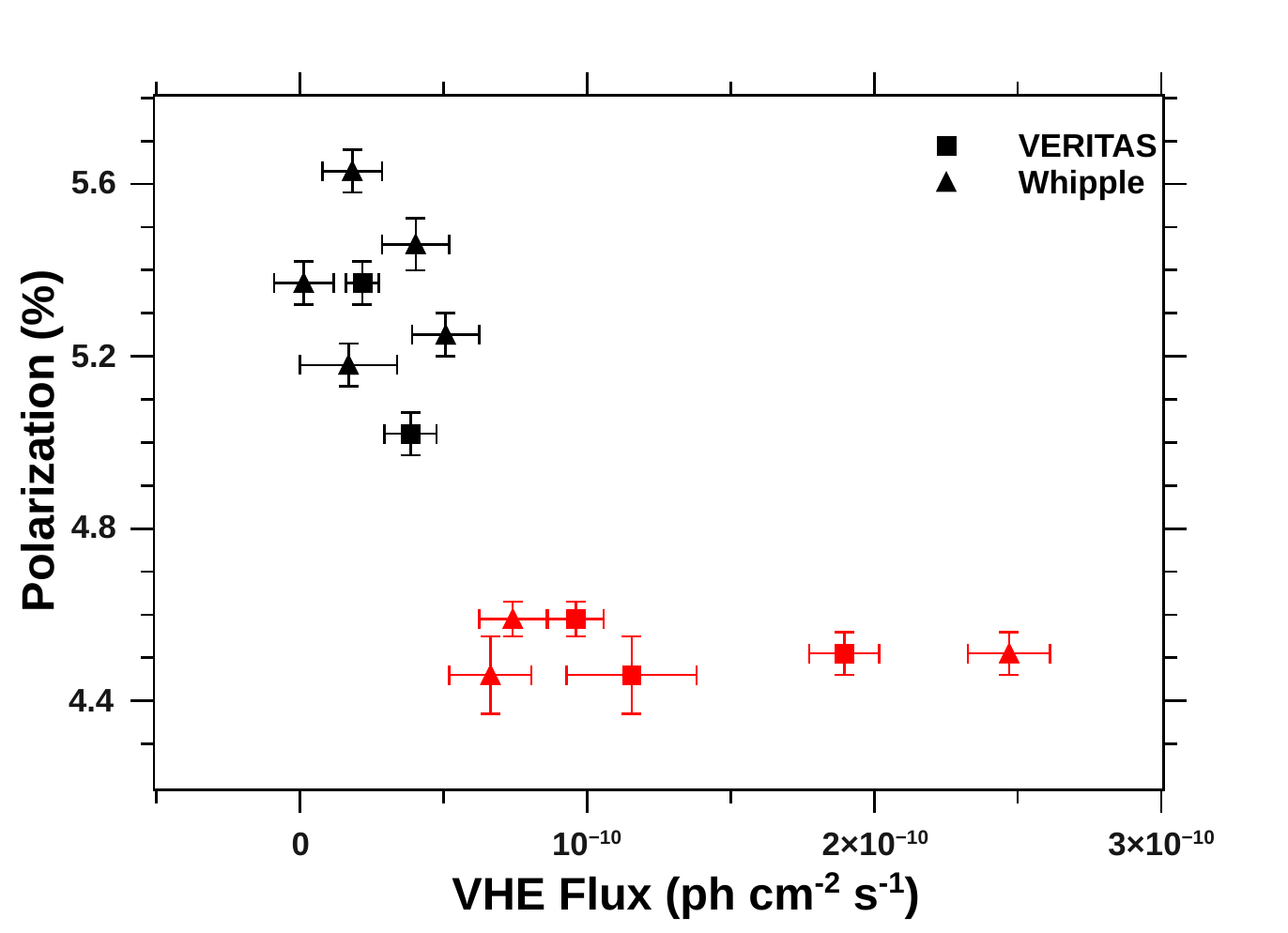}
  \includegraphics[width=0.97\hsize]{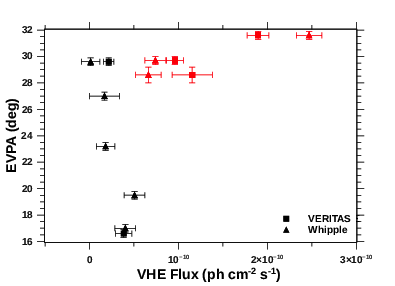}
  \caption{Correlation of VHE $\gamma$ rays with optical polarization ({\it top}) and EVPA ({\it bottom}), in which only pairs of observations within 24 hours of each other were used. The events taken with Whipple (filled triangles) and VERITAS (filled squares) before (black) the flare are well differentiated from the events after (red) the VHE flare.}
  \label{corr-pol-vhe}
 \end{figure}

Optical flux observations showed no correlation with VHE measurements. 
A comparision of the optical polarization and VHE light curves given in Figure \ref{m5pol} shows evidence for a correlation which seems to be present in coincidence with the flare.
To better appreciate this behavior, Figure \ref{corr-pol-vhe} shows the degree of optical linear polarization (top panel) and the EVPA (bottom panel)  plotted against the VHE flux taken with both Whipple and VERITAS. Again, a coincidence window of 24 hours was considered. The degree of polarization is different for observations taken before (black points) and after (red points) the VHE flare, clearly showing a 15\% drop from 5.3\% to 4.5\% after the VHE flare, mentioned in Section \ref{lc}. The EVPA plot shows that the increase occurred at approximately constant VHE flux before the flare, and remained at the highest values during the high flux stage of the source at VHE. 

\begin{figure}[!h]
  \vspace{5mm}
  \centering
\includegraphics[width=3.5in]{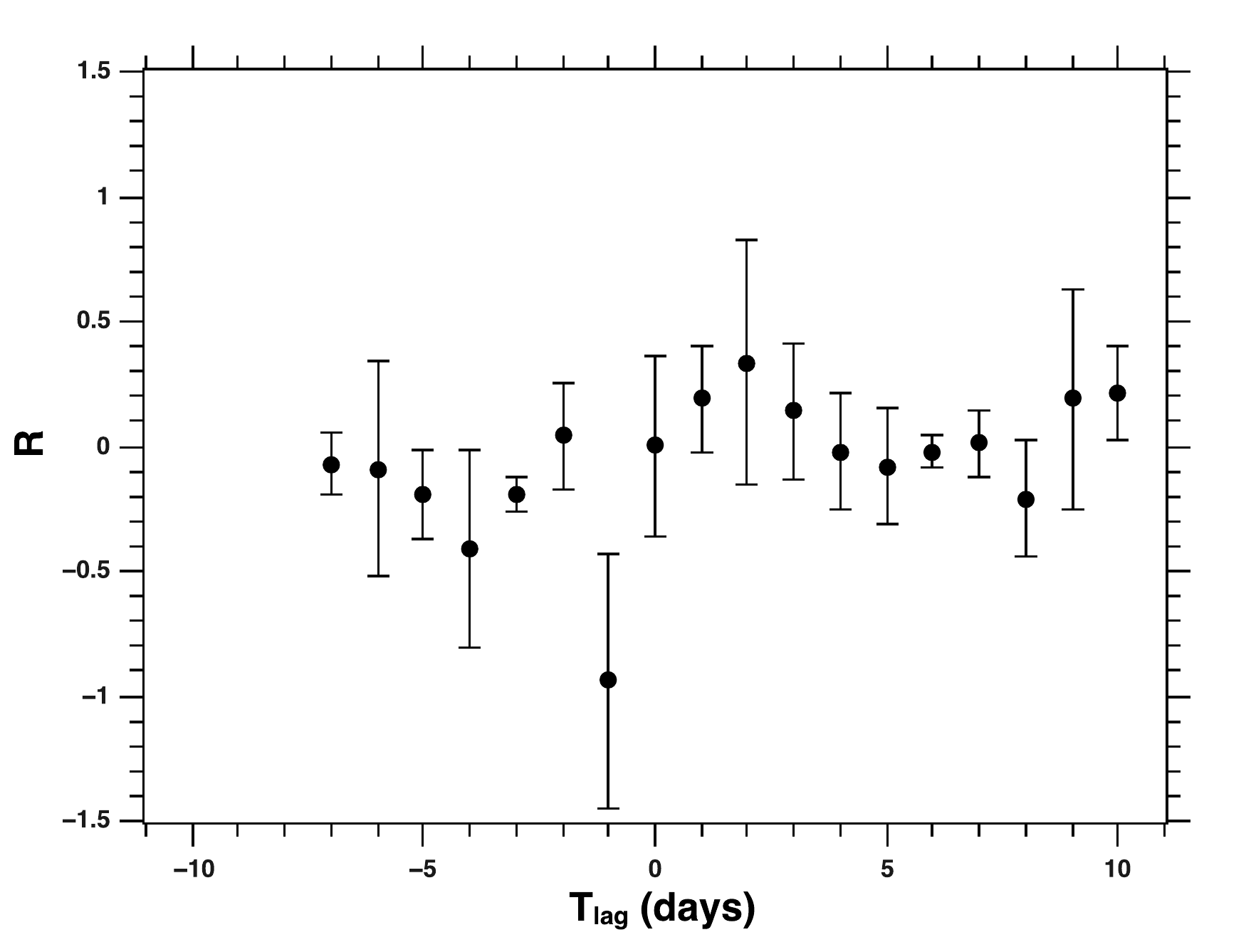}
  \caption{Discrete correlation function (DCF) of the VHE gamma-ray
light curve with respect to the X-ray light curve.}
  \label{dcf}
 \end{figure}

The discrete correlation function (DCF) as outlined in \citet{EK88} was also computed to search for correlations between discrete emission measurements in the VHE and X-ray bands at several time lags. This method is an approximation of the standard correlation function that works with functions not well-sampled and with data points with statistical uncertainties of the same order of magnitude as the flux variations, as is the case for the light curves used in this work. Figure \ref{dcf} shows the DCF values in steps of 1-day, for time lags covering the days of the X-ray observations.
No significant correlations were found. 
This result does not exclude the existence of an X-ray flare, including one that is substantially smaller than that seen at VHE, as occurred for PKS\,2155$-$304 in 2006 \citep{lab20}. This would be consistent with the intrinsic variations calculated from the {\it Swift}-XRT, data mentioned in Section \ref{lc}. \\

\section{Spectral energy distribution}
\label{SED}

\begin{figure}
  \centering
\includegraphics[width=\hsize]{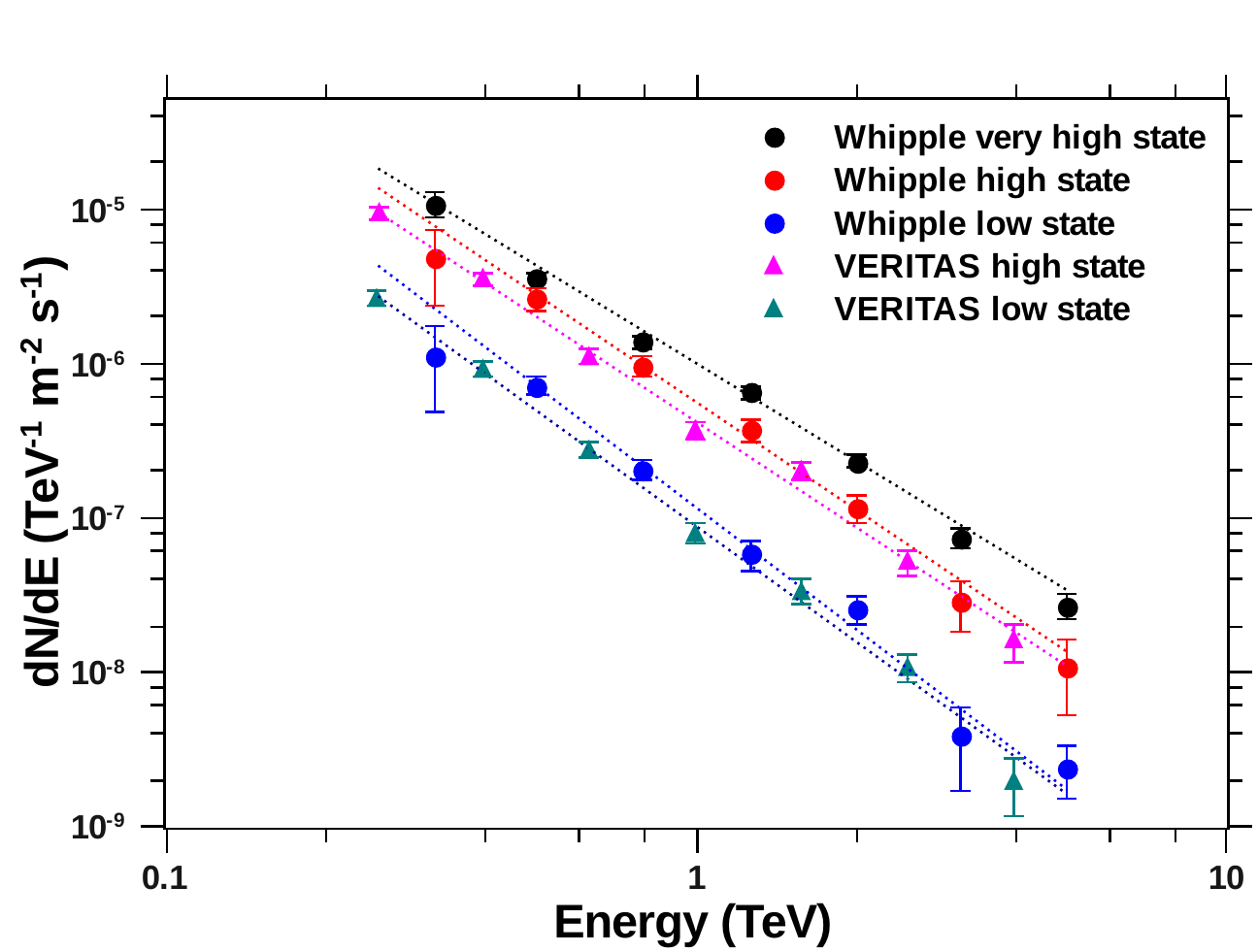}
  \caption{Time-averaged VERITAS and Whipple spectra of Mrk 501 for discrete flux levels (see text).}
  \label{fig5}
 \end{figure}

\begin{figure}
  \centering
\includegraphics[width=\hsize]{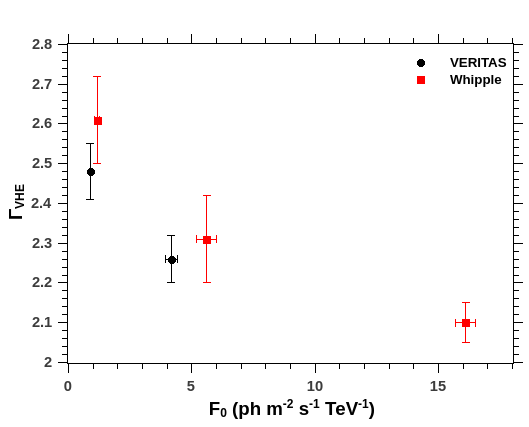}
  \caption[]{VHE Photon index vs. flux normalization ($F_0$) obtained from the power-law fits of Figure \ref{fig5} for the different periods of activity as defined in Table \ref{table_single}. A linear fit was done to all the data, obtaining a $\chi^{2}$ of 8.2 with a p-value of 0.09.}
  \label{figphotonindex}
 \end{figure}

The study of the spectral evolution of blazars is important for the understanding of the acceleration mechanism in the jets, particularly during flares. Blazars have shown spectral variability dependent on the flux-level, with, in some cases, a clear hardening when the flux level increases. The differential energy spectra of Mrk\,501 for VHE $\gamma$ rays are shown in Figure \ref{fig5} for the Whipple and VERITAS observations taken in the period reported here. They were modeled in each case, for the quiescent emission and the flaring state, with a simple power law 
${dN}/{dE}= F_{\rm 0} {\times 10^{-7}}~ (E/1~\rm TeV)^{-\Gamma_{ \rm VHE}} ~~ {\rm ph~ m^{-2}~ s^{-1}~ TeV^{-1}}$
, where $F_{\rm 0}$ is a normalization factor and $\Gamma_{ \rm VHE}$ is the photon index.
The best fit for each set is also shown in Figure \ref{fig5}, and the parameters and associated errors are summarized in Table \ref{table_single}. An indication of spectral hardening with increasing flux activity was found for the TeV band, as shown in Figure \ref{figphotonindex}. The softest photon index was 2.61$\pm$0.11 (for the low/medium state), and the hardest photon index was 2.10$\pm$0.05 (for the flare on MJD 54952). 
A similar trend had already been found during 2005 with MAGIC \citep{alb07}, having 2.45$\pm$0.07 and 2.28$\pm$0.05 for the low and high state, respectively. 

\begin{table*}
\caption{Best-fit parameters for VHE spectra at different flux states, as shown in Figure \ref{fig5} (${dN}/{dE}= F_{\rm 0} {\times 10^{-7}}~ (E/1~\rm TeV)^{-\Gamma_{ \rm VHE}} ~~ {\rm ph~ m^{-2}~ s^{-1}~ TeV^{-1}}$).}             
\label{table_single}      
\centering 
\begin{tabular}{c c c c c}
\hline
 & MJD Interval &  $F_{\rm 0}{\times 10^{-7}}~{\rm ph~ m^{-2}~ s^{-1}~ TeV^{-1}}$      &  $\Gamma_{ \rm VHE}$ & $\chi^{2}$/NDF \\
\hline
Whipple very high  & 54952.35-54952.41 & 16.1 $\pm$ 0.4    & 2.10 $\pm$ 0.05  & 13.48/8    \\
Whipple high       & 54952.41-54955    & 5.60 $\pm$ 0.40     & 2.31 $\pm$ 0.11  & 3.10/8     \\
Whipple low        & 54936-54951       & 1.16 $\pm$ 0.09   & 2.61 $\pm$ 0.11  & 3.40/8     \\
VERITAS high       & 54952-54955       & 4.17 $\pm$ 0.24   & 2.26 $\pm$ 0.06  & 6.26/5     \\
VERITAS low        & 54938-54951       & 0.88 $\pm$ 0.01  & 2.48 $\pm$ 0.07  & 3.76/5     \\
\hline
\end{tabular}

\end{table*}

The spectral analysis of the {\it Fermi}-LAT observations was performed for two different time periods based on the VHE flux. The analysis of the first period, from April 17 to April 30, 2009 (MJD 54938--51) results in a significant detection at a level of 5.7$\sigma$. The spectrum is well described with a power law fit with F(E$>300$ MeV)=(1.1$\pm$0.4)$\times10^{-7}~{\rm erg~cm^{-2}~s^{-1}}$ and spectral index $\Gamma$=2.0$\pm$0.2. The second integration period corresponds to May 1 to 5, 2009 (MJD 54952--56) with a significance of 6.4$\sigma$. The spectrum is also compatible with a power-law with F(E$>300$ MeV)=(5.0$\pm$2.9)$\times10^{-8}~{\rm erg~cm^{-2}~s^{-1}}$ and spectral index $\Gamma$=1.6$\pm$0.3.

The X-ray spectral analysis was also performed for the two different time periods based on the VHE flux. For the first state (low state), the spectrum is well described with a log-parabola ${\rm F(E)=K \cdot (E/keV)^{(-\alpha - \beta \cdot log(E/keV)}}$ fit with K=(2.51$\pm$0.04) $\times10^{-2}~{\rm ph~cm^{-2}~s^{-1}~keV^{-1}}$, $\alpha$=1.84$\pm$0.03, $\beta$=0.17$\pm$0.07 and a $\chi^{2}$/dof=8.57/7. For the second state (high state), the spectrum is also well described with a log-parabola fit with K=(2.97$\pm$0.02)$\times10^{-2}~{\rm ph~cm^{-2}~s^{-1}~keV^{-1}}$, $\alpha$=1.81$\pm$0.28, $\beta$=0.166$\pm$0.071 and a  $\chi^2$/dof=6.13/7.

SED modeling was carried out by using a pure SSC model, based on \citet{BC02}. The equilibrium version of the model is described more thoroughly in \citet{Bot13}.  
In this one-zone model, a power-law energy distribution of electrons of the form $Q(\gamma) = Q_{0} \gamma^{-q}$ between a minimum energy $\gamma_{min}$ and a maximum energy $\gamma_{max}$, is injected into the emission region. The radiation code then evaluates self-consistently an equilibrium between this injection, radiative cooling, and particle escape on a timescale $t_{esc} = \eta R/c$, in which $\eta > 1$ is the escape timescale parameter, resulting in a broken power-law equilibrium distribution.
The emitting region at the comoving radius ${\it R_{B}}$ moves along the jet with a relativistic speed $\beta$. The particles cool due to radiative losses and then might escape from the region. The viewing angle $\theta$, between the jet direction and the line of sight, is set to be the superluminal angle, where the bulk Lorentz factor $\Gamma$ equals the Doppler factor. The values for the parameters of the model are shown in Table \ref{sed}. 

The model includes only synchrotron and inverse-Compton (IC) emission, since this is the model with the fewest free parameters, and it is usually sufficient to fit the SEDs of HBLs like Mrk\,501. In particular, such a model provides a satisfactory fit to the SEDs of Mrk\,501 presented here. For example, bremsstrahlung is included in the B\"ottcher et al. (2013) code, but is generally insignificant relative to synchrotron and IC contributions for the density parameters required to model the X-ray and $\gamma$-ray fluxes.
We note in Table \ref{sed} that, for both states, magnetic fields far below equipartition are required. 
This behavior was also found previously for Mrk\,501 \citep{man12} and for other TeV blazars (e.g. 1ES\,1312-432 \citep{abra13b}, Mrk\,421 \citep{abdo11d} and SHBL\,J001355.9-185406 \citep{abra13}). The low field is required to facilitate slight energetic dominance of the inverse Compton component, and to prevent the synchrotron peak from moving to the hard X-ray energies seen in the 1997 flare of Mrk 501 (Acciari et al. 2011a).

The transition between the two states could not be achieved by changing only one or two parameters.
For the size of the emission region to remain the same, and considering that the dimensions of the jet are unlikely to change much in a few days (only constrained by allowing at least for intra-day variability), it was necessary to change the Doppler factor, the high-energy cutoff of the injected electron distribution ($\gamma_{max}$), the electron injection index {\it q}, and the magnetic field. Changes in $\gamma_{max}$ usually reflect variations in the radiative cooling rate, and changes in $q$ signify modifications of the turbulent acceleration environment.  Subtleties pertaining to these are discussed at length in \citet{baring16}, where, in particular, lower values of $q\sim 1$ are used in modeling Mrk 501; such flat distributions correspond to cases where turbulence levels are low enough to permit the action of coherent drift acceleration in jet shocks. This inference of low field turbulence may have significant implications for the optical polarization observations, since it is consistent with significant coherence of fields on large scales.  We note that the value of $q$ is poorly constrained by the {\it Fermi}-LAT data, particularly since the obvious steepening into the VERITAS band probably begins in the upper end of the LAT energy range.
To account for the optical emission of the host galaxy, a thermal blackbody core with a temperature of 10 000\,K \citep{rous11} was added to fit this set of data, giving a much better overall fit; this portion of the broadband spectrum is extremely difficult to model with a synchrotron component. Furthermore, 
in this work, the UVOT data can be accounted for as part of the host galaxy, although it is not well established what the origin is as the UV contribution
could be due to the host galaxy emission or the synchrotron emission (e.g \citet{abdo10}).

\begin{figure}
  \centering
\includegraphics[width=\hsize]{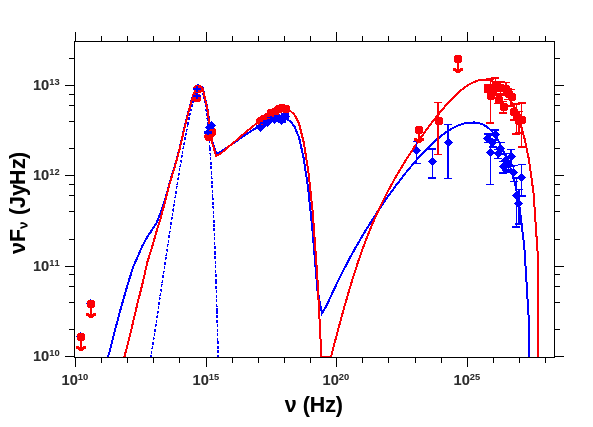}
  \caption[]{Spectral energy distribution of Mrk\,501 for the low state (MJD 54936--54951; blue squares) and high state (MJD 54952--55; red circles) of the 3-week period. The SSC model representative for low (blue solid line) and high (red solid line) states is also shown. The blue dotted line corresponds to the optical emission of the host galaxy.}\label{m5-sed}
 \label{m5_sed}
 \end{figure}

The SED of Mrk\,501 for the low state (MJD 54936--54951) and high state (MJD 54952--55, including the very high) are shown in Figure \ref{m5-sed}, together with the results from the SSC model that is representative of both states.
It can be seen from the figure that the more significant spectral variability was seen at the highest energies of the spectral energy distribution. 

\begin{table}
\caption{SED model parameters for the low state (MJD 54936--54951) and high state (MJD 54952--55).}
\label{sed}
\centering
\begin{tabular}{c c c }
\hline
 Parameters &  Low state &  High state  \\
\hline
$\gamma_{min}$ & $1 {\times 10^{4}}$ & $1 {\times 10^{4}}$ \\
$\gamma_{max}$ & $1.2 {\times 10^{6}}$ & $2 {\times 10^{6}}$ \\
Injection electron spectral index ({\it q}) & 1.6  & 1.5 \\
Escape time parameter ($t_{esc}=\eta R/c$) & 1000 & 1000  \\
Magnetic field [G] & 0.03 & 0.0075   \\
Blob radius (${\it R_{B}}$) [cm]  & $1.2 {\times 10^{16}}$  & $1.2 {\times 10^{16}}$  \\
Electron power ($L_{e}$) [erg $s^{-1}$] & $8.27 {\times 10^{43}}$ & $2.53 {\times 10^{44}}$     \\
Poynting Flux ($L_{B}$) [erg $s^{-1}$] & $1.94 {\times 10^{41}}$    & $2.73 {\times 10^{40}}$    \\
$L_{B}/L_{e}$  & $2.35 {\times 10^{-3}}$  & $1.08 {\times 10^{-4}}$    \\
\hline
\end{tabular}

\end{table}

\section{Summary}

Multi-wavelength observations of Mrk\,501 were undertaken from April 17 to May 5, 2009 with a number of ground- and
space-based observatories covering the electromagnetic spectrum from radio to VHE $\gamma$ rays. 
The main purpose of this work is to analyze the VHE flare of May 1, which was first
detected by the Whipple 10\,m $\gamma$-ray telescope, and its correlation with other bands, to help identify the processes involved during this emission.

Light curves were analyzed for all wavebands involved. At VHE, the light curve was consistent with constant emission of the source until the night of May 1
(MJD 54952), when a high-emission state was detected first with Whipple and later with
VERITAS, reaching a maximum $\gamma$-ray flux of $\sim$10 times the average baseline flux (approximately five times the Crab Nebula flux), and showing an increase of a factor $\sim$4 in  25 minutes.
The fluxes measured at lower energies did not show any significant variation before or after the VHE flare, except for the {\it Swift}-XRT and UVOT fluxes, which exhibited moderate variability.

The optical polarization and the polarization position angle both show unusual polarization behavior, as compared to observations of this source in the past, reaching a level of $\sim5.6\%$, one of the highest levels observed. The EVPA reached $\sim$30 degrees at the peak of the $\gamma$-ray flare and reversed its direction of rotation as the outburst faded. These measurements seem to correlate with the VHE flare, indicating a possible common origin, as has occurred for several other outbursts reported recently for other sources. 
The correlation between optical polarization and EVPA with a VHE flare is not very common and it was only observed in a couple of HBLs before, like PKS\,1510$-$089 \citep{lab17}. This is the first observation of this behavior displayed by Mrk\,501.
Studying the correlation between VHE activity and polarization changes could be a good opportunity to find a new scenario for the VHE flares and could be an alternative method to predict them. Many of the MW campaigns conducted have some observations of the optical polarization, but so far the flares detected at X-ray and $\gamma$-ray energies do not have any quasi-simultaneous polarization observations. Therefore, it is important to conduct a long-term campaign covering the optical polarization and the $\gamma$-ray band before making conclusions about the correlation between them. There are some new experiments dedicated to studying the connection between rotations in the optical polarization and flares in the $\gamma$-ray band, like RoboPol \citep{bli}.

The differential energy spectra for the VHE $\gamma$ rays were calculated for the Whipple and VERITAS observations taken in the period reported here. They were modeled with a simple power law for all the states. 
An indication of spectral hardening with increasing flux activity was found, with the softest photon index in the range 2.5-2.6 in the low state, 2.25-2.3 in the high state, and 2.10$\pm$0.05 for the highest state during the flare on May 1. 
SED modeling was carried out for the quiescent (MJD 54936--51) and high states (MJD 54952--55; including the high and very high state from Table \ref{table_single}) of activity, by using a pure SSC model from which a set of parameters was found. The main differences between these parameters and those found in the overall campaign \citep{abdo10} is that, in the present study, the Doppler factor was bigger (20-30 compared to 12 obtained using the main SSC fit), the magnetic field was even lower (0.03-0.0075 G compared to 0.015\,G), and the emission region was smaller ($\sim1.2 \times 10^{16}$ cm). However, the parameters obtained here are in concordance with the alternative SSC fit used also for the overall campaign (see \citet{abdo10} for further information).

It is clear from the data taken during this campaign that fast flaring activity has been detected in the band of VHE $\gamma$-rays, with fall and rise times of the order of a few tens of minutes. Moderate spectral variability was also observed in the spectra at different states of flux activity. 
Given that this source is known to exhibit very fast variability, the fact that we do not have strictly simultaneous observations between X-rays and VHE $\gamma$ rays makes it difficult to draw firm conclusions about correlations.

\begin{acknowledgements}
     This research is supported by grants from the U.S. Department of Energy Office of Science, the U.S. National Science Foundation and the Smithsonian Institution, and by NSERC in Canada. We acknowledge the excellent work of the technical support staff at the Fred Lawrence Whipple Observatory and at the collaborating institutions in the construction and operation of the instrument. The VERITAS Collaboration is grateful to Trevor Weekes for his seminal contributions and
leadership in the field of VHE gamma-ray astrophysics, which made this study possible.
The Fermi LAT Collaboration acknowledges support from a number of agencies and institutes for both development and the operation of the LAT as well as scientific data analysis. These include NASA and DOE in the United States, CEA/Irfu and IN2P3/CNRS in France, ASI and INFN in Italy, MEXT, KEK, and JAXA in Japan, and the K. A. Wallenberg Foundation, the Swedish Research Council and the National Space Board in Sweden. Additional support from INAF in Italy and CNES in France for science analysis during the operations phase is also gratefully acknowledged.
MB acknowledges support through the South African Research Chairs Initiative (SARChI) by the National Research Foundation and the Department of Science and Technology of South Africa.
\end{acknowledgements}

\end{document}